\documentclass[preprint,12pt]{elsarticle}

\usepackage{xcolor}
\usepackage{graphicx}
\usepackage{dcolumn}
\usepackage{bm}
\usepackage{float}
\usepackage{amsmath}
\usepackage{siunitx}

\usepackage[left=2.5cm,right=2.5cm,top=2cm,bottom=1.5cm,includeheadfoot]{geometry}

\begin{document}

\begin{frontmatter}
\title{Implementation of Light Diagnostics for Wakefields at AWAKE}

\author[1,2]{J. Mezger}
\author[1,3]{M. Bergamaschi}
\author[1]{L. Ranc}
\author[3]{A. Sublet}
\author[1,3]{J. Pucek}
\author[3]{M. Turner}
\author[1,2]{A.~Clairembaud}
\author[1]{P. Muggli}

\affiliation[1]{
organization={Max-Planck-Institute for Physics},
city = {Munich},
country = {Germany}
}
\affiliation[2]{
organization={Technical University of Munich},
city = {Munich},
country = {Germany}
}
\affiliation[3]{
organization={CERN},
city = {Geneva},
country = {Switzerland}
}

\begin{abstract}
We describe the implementation of light diagnostics for studying the self-modulation instability of a long relativistic proton bunch in a 10\,m-long %
plasma. %
The wakefields driven by the proton bunch dissipate their energy in the surrounding plasma. %
The amount of light emitted as atomic line radiation is related to the amount of energy dissipated in the plasma. %
We describe the setup and calibration of the light diagnostics, configured for a discharge plasma source and a vapor plasma source. %
For both sources, we analyze measurements of the light from the plasma only (no proton bunch). %
We show that with the vapor plasma source, the light signal is proportional to the energy deposited in the vapor/plasma by the ionizing laser pulse. We use this dependency to obtain the parameters of an imposed plasma density step. %
This dependency also forms the basis for ongoing studies, focused on investigating the wakefield evolution along the plasma. %

\end{abstract}

\end{frontmatter}

\section{Introduction}
A charged particle bunch propagating through plasma drives wakefields by displacing plasma electrons, thus exciting a plasma oscillation that travels with the drive bunch~\cite{chen1985acceleration}. %
For effective driving of wakefields, the length of the drive bunch needs to be on the same order as the wavelength of the plasma oscillation. %
The distance over which the wakefields can be driven depends on the energy carried by the particles and by the bunch. %
With electron bunches, very high charge densities can be reached with the bunch lengths required for effective wakefield excitation. %
Therefore, wakefields can be effectively driven by a single, short bunch. %
This was demonstrated at SLAC by generating an accelerating gradient of 50 GeV/m~\cite{SLAC42GeV}. %
Because of the energy carried by the 
electron bunch, this gradient could be driven over only 85\,cm. %
By utilizing a proton bunch carrying a much higher energy, the acceleration could be driven over a longer distance. %
However, proton bunches are only available with relatively low charge density and are many times longer than the plasma wavelength at plasma densities required to drive large amplitude wakefields. %
Such bunches can drive wakefields with amplitudes in the GV/m range, only after undergoing self-modulation~\cite{kumar2010self}. %
The self-modulation instability grows from initial wakefileds in the MV/m range. %
The focusing and defocusing fields of these wakefields modulate the bunch density, eventually transforming the long bunch into a train of micro-bunches. %
The spacing between micro-bunches is approximately equal to the period of the wakefields. %
Each micro-bunch drives wakefields, coherently adding to the wakefields driven by the previous ones. %
Therefore, as the micro-bunch train develops, the wakefield amplitude grows along the bunch. %
Numerical simulation results show that, for plasma densities around $n_{pe} = 7\times10^{14}$\,cm$^{-3}$ and typical parameters of the incoming bunch of this experiment, a fully developed micro-bunch train can drive wakefields with amplitudes on the order of GV/m~\cite{lotov2014parameter}. %

AWAKE has demonstrated acceleration of electrons from 19\,MeV to 2\,GeV in a 10\,m-long plasma~\cite{AWAKEnature}. %
It is not possible to determine the accelerating gradient from these measurements, because acceleration mainly happens after the micro-bunch train is fully developed, meaning over a shorter, and currently unknown distance. %
It is therefore of great interest to establish a diagnostic that can measure the development of the wakefields as self-modulation occurs~\cite{AAC2022_plasmalight}. %
The light diagnostic relies on the assumption that the energy of the wakefields is locally dissipated in the plasma. %
The energy dissipates through many different processes, including the collisional excitation of bound electrons in the shell of neutral atoms and ions, as well as through the recombination of free electrons and ions. %
These processes lead to the emission of light at the wavelength of the spectral lines of the atoms. %
Therefore, one can expect the amount of light measured by the diagnostic to scale with the energy deposited in the plasma by the drive bunch, as was observed in reference~\cite{SLAC-plasmalight}.

In this paper we present the implementation of light diagnostics for a vapor plasma source (VPS) and a discharge plasma source (DPS). %
The newly commissioned VPS features ten view ports along the ten meters of the source~\cite{AAC2022_plasmalight}. %
At each view port location, the light is recorded by a CMOS camera and a photomultiplier tube (PMT). %
In the DPS, the plasma is created in a glass discharge tube. %
Therefore, it is possible to capture images of large, uninterrupted sections of the plasma, over its whole length. %
Two CMOS cameras record the spatial distribution of light along the plasma. %
In the following, we discuss the setup and the calibration of the instruments used. %
We also analyze measurements with the discharge and vapor plasmas, without propagating a proton bunch (no wakefields). %
For the vapor plasma, we analyze the experimental data with the aid of a simple model for the energy deposited in the plasma by the ionizing laser pulse, similarly to what is shown in reference~\cite{oz2014VAPS}. 

\section{Implementation of the Diagnostic}

    \subsection{Discharge Plasma Source} \label{sec:DPS-setup}
\begin{figure}[t]
        \centering
        \includegraphics[width=1\textwidth]{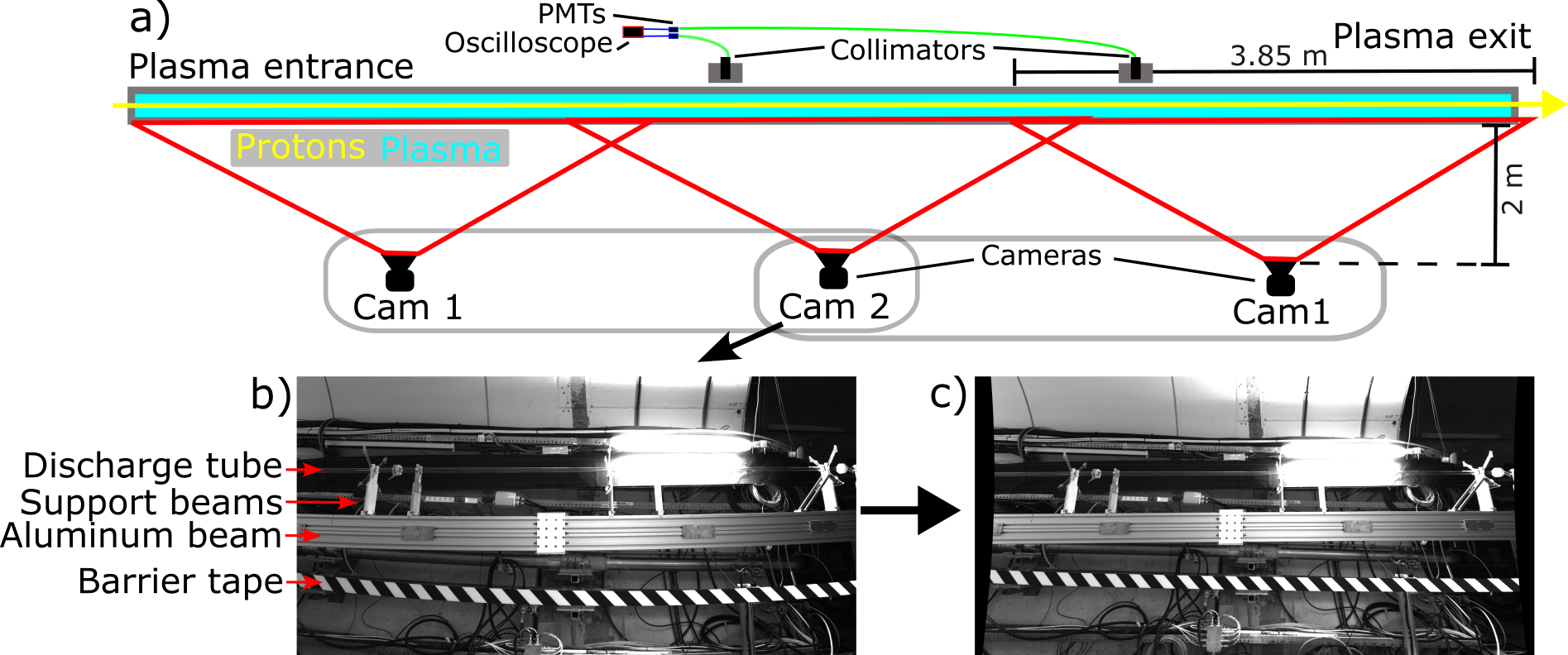}
        \caption{(a) Schematic drawing of the experimental setup for the DPS. %
        Proton bunches propagate from left to right. %
        (b-c) Reference image of camera 2 in the experiment (b) before, and (c) after correcting for barrel distortion and vignetting. %
        }
        \label{fig:DPS-setup}
    \end{figure}

    For the DPS, the diagnostic is composed of two CMOS cameras with wide angle lenses for a 84$^\circ$ field of view in the horizontal plane. %
    Because of the limited space available in the experiment, cameras are placed at a distance of only two meters from the source. %
    Therefore, each camera covers a 3.85\,m section of the plasma. %
    Allowing for some overlap, they have a combined field of view of seven meters (see Fig.~\ref{fig:DPS-setup} a)). %
    Because in most cases with protons, the amount of light emitted by wakefields in the first three meters of the plasma (not shown) is below detection threshold, the two cameras are set to record light from the last seven meters of the plasma. %
    However, we present here some results for the whole ten meters of plasma obtained by temporarily moving the second camera to cover the first three meters of the source (all other parameters equal). %
    
    Images captured by the cameras have to be corrected for optical and geometrical effects that influence the amount of light recorded by each pixel along the plasma, in the horizontal plane of the image. %
    We correct the images for barrel distortion, vignetting, and consider the angle at which the plasma is observed ($\pm$ 42°). %

    Barrel distortion is introduced by the wide angle lenses (lens: Fujinon $f = 8$\,mm, sensor: Sony IMX 530 14.6~(h)\,mm $\times$ 12.6~(v)\,mm) and causes pixels further from the image center to cover a larger area in real space than pixels closer to the image center. %
    To correct for this distortion, we use Zhang's method \cite{Zhang}. %
    Once the camera matrix and distortion coefficients are calculated, they can be used to map a grid of pixels in a corrected image to the respective positions in the raw image. %
    We use an area-based interpolation method to calculate the pixel counts in the corrected image, based on the area they cover in the raw one. %
    This procedure conserves the number of counts, i.e. the amount of light recorded. %
    The pixels in the corrected image display the counts recorded over equal areas, regardless of their distance to the image center. %
    After correction, the area along the plasma covered by each pixel is $\sim$0.8$\times$0.8\,mm$^2$ (discharge tube outer diameter 30 mm, inner diameter 26 mm). %
    In Fig.~\ref{fig:DPS-setup} b)-c) we see that straight lines such as those of the barrier tape or the aluminum beam only appear as straight in the image after correcting for distortion (c)). %
    
    To correct for vignetting, we record images of an evenly illuminated area. %
    We sum pixel counts in the vertical direction over the area where the plasma would be located in the image to obtain the vignetting function. 
    Pixel counts recorded by the camera are divided by the vignetting function. %
    
    Because the plasma is optically thin and cylindrical, light from entire cross sections reaches the camera. %
    Therefore, the amount of light reaching the image sensor also depends on the angle at which the plasma is observed. %
    Observing the plasma at an angle $\alpha$ increases the area of the cross section by a factor $\frac{1}{\cos{\left( \alpha \right)}}$. %
    The amount of light detected at an angle thus increases with the observation angle. %
    We correct for this effect by multiplying the pixel values recorded by the camera by $\cos{\left( \alpha \right)}$. %
    
    The exposure time of the cameras is set to 1\,\textmu s. %
    This value is chosen because PMT signals show that the emission of light due to wakefield dissipation (not shown) only lasts for several hundred nanoseconds in this case. %
    To capture the temporal evolution of the light, two PMTs are used. %
    Light is coupled into optical fibers through parallel collimators at two locations along the plasma, and transmitted to the PMTs. %
    The collimators are mounted 4.5\,m and 6.9\,m from the plasma entrance. 
    
    
    \begin{figure}[!b]
        \centering
        \includegraphics[width=1\textwidth]{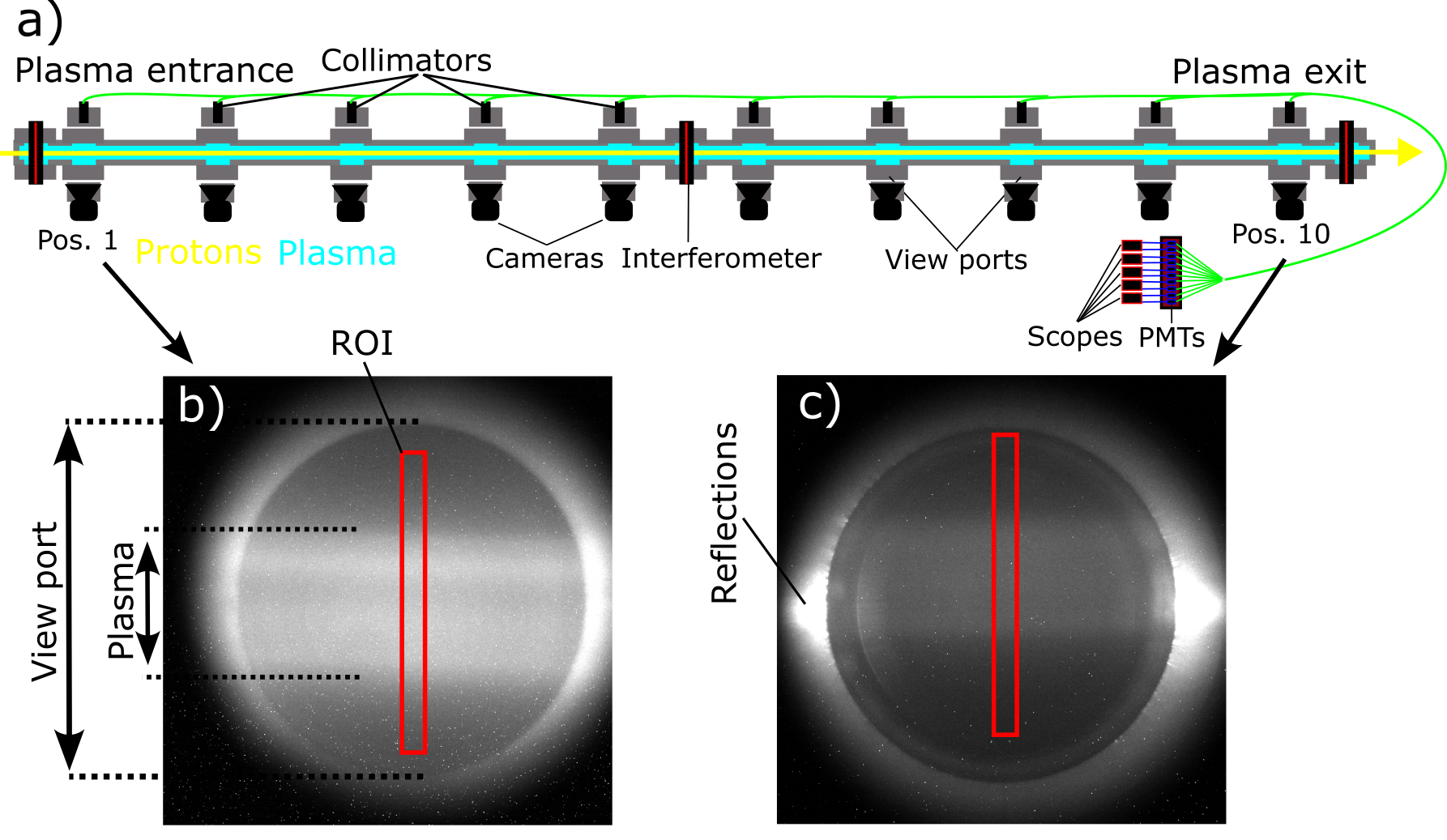}
        \caption{(a) Schematic drawing of the setup for the VPS. (b-c) Images of the plasma at view port position (b) 0.5\,m and (c) 9.5\,m. %
        Red rectangles indicate ROIs used for calculating the light signal. %
        View port diameter 15\,mm, ROI width 1\,mm, height 13\,mm. %
        Same maximum counts values for both images. %
        }
        \label{fig:VAPS-setup}
    \end{figure}
    
    \subsection{Vapor Plasma Source}

    For the VPS, the light diagnostic is limited to recording signals at ten view port positions along the plasma. %
    At each position, we place a CMOS camera and a fiber collimator for the PMTs (Fig.~\ref{fig:VAPS-setup} a)). %
    We use an exposure time for the cameras between 40\,\textmu s and 150\,\textmu s. %
    This is because the signal from wakefield dissipation (not shown) lasts for much longer, compared to the DPS. %

    For the measurements to be comparable between PMTs and cameras, and between the different view port positions, the fields of view of the collimators and cameras have to be aligned to the plasma. %
    For the cameras, the live image can be used to center the camera on the view ports (i.e. also on the plasma). %
    For the PMTs, a white light source is used to couple light into an optical fiber. %
    This light is then coupled out by the collimators such that the optical axis of the collimators is made visible and can be aligned to the location of the plasma in the viewport. %
    The camera images are used to observe the transverse distribution of plasma light and to assess the width of the plasma column. %
    We note here that no correction is applied to these images since we use a standard lens with focus at a short distance ($\sim$30\,cm). %
    To obtain a light signal, we sum the counts over a strip in the center of the view port (see Fig.~\ref{fig:VAPS-setup} b)-c)). %
    This is necessary to avoid effects on the sides of the viewport window, such as reflections, or changes in the relative volume of the plasma column that is seen through the circular view port window, and caused by changes in the transverse size of the column (cross section between two cylinders). %
    The viewport diameter is 15\,mm (pixel size $\sim$0.025$\times$0.025\,mm$^2$). %
    We choose the region of interest (ROI) to be 1\,mm in width and 13\,mm in height, as indicated in Fig.~\ref{fig:VAPS-setup} by the red rectangles. %

\section{Plasma}

    \subsection{Discharge Plasma}
\begin{figure}[t]
        \centering
        \includegraphics[width=1\textwidth]{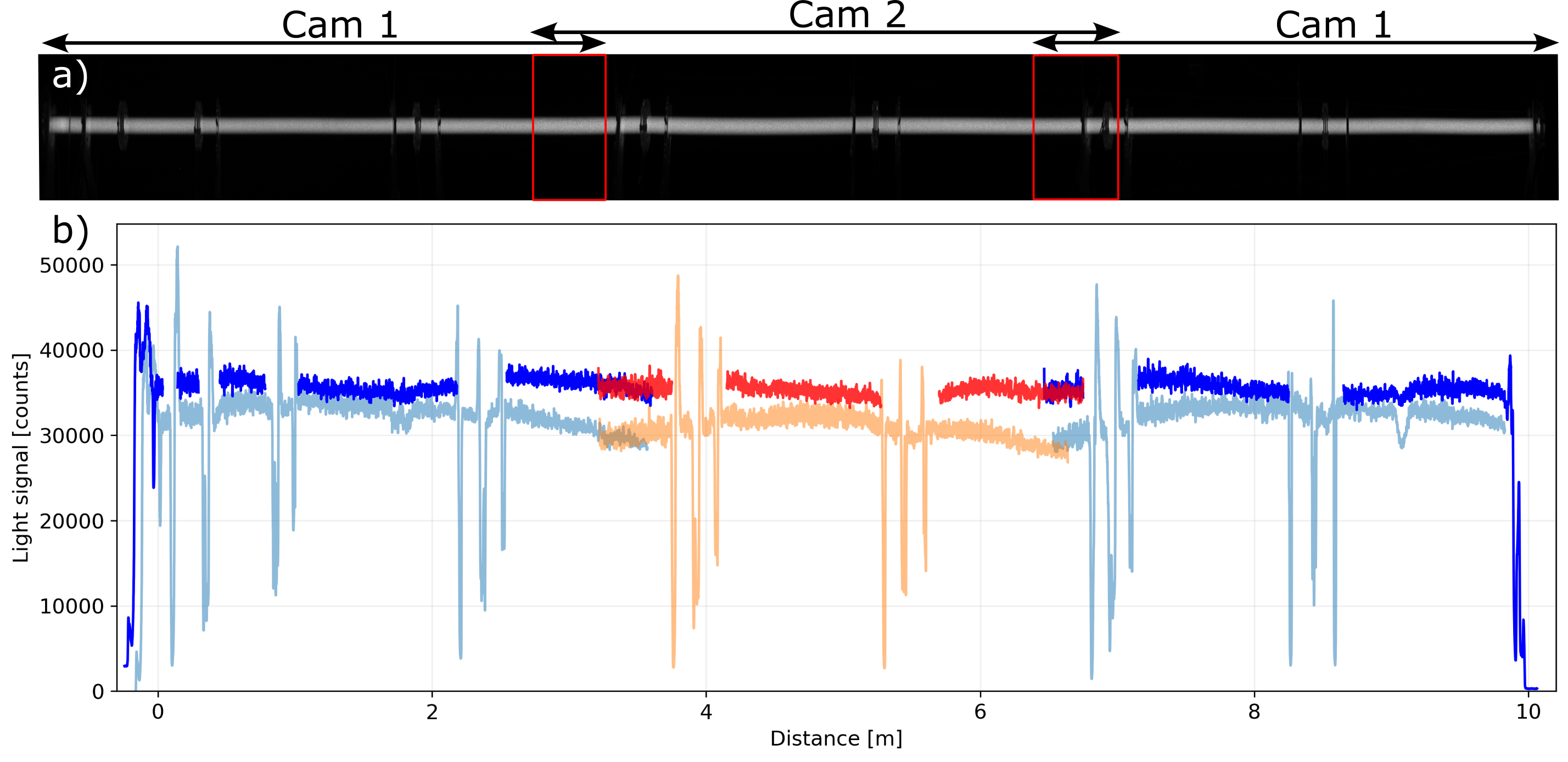}
        \caption{(a) Images of the discharge plasma %
        with the overlapping regions marked by red rectangles. %
        Image on the left (Cam 1) obtained on a different day, with the same plasma parameters (24\,Pa of Ar, 500\,A, 80\,\textmu s delay, $n_{pe} = \SI{4.8e14}{cm^{-3}}$) as the middle (Cam 2) and right (Cam 1) images. 
        (b) Plasma light profiles (distance measured from the plasma entrance), taken with camera 1 (blue) and camera 2 (red). %
        Profiles from uncorrected images are plotted with light colors. %
        Vertical stripes in the uncorrected profiles caused by support beams of the source are masked in the corrected profiles. %
        Uncorrected profiles are offset by -4000 counts for better visibility. %
        Profiles show eight events plotted on top of each other. }
        \label{fig:DPS-cams}
    \end{figure}

    In the DPS, the plasma is created by drawing a current 
    through the discharge tube~\cite{NunoDPS,Carolina_DPS}. %
    The current pulse, with a duration of 30\,\textmu s, creates a plasma with peak electron densities $n_{pe}$ between $\SI{3e14}{cm^{-3}}$ and $\SI{19e14}{cm^{-3}}$, depending on the discharge current, the type of gas in the tube, and its pressure. %
    After the current pulse, ions and electrons recombine. %
    By delaying the arrival time of the proton bunch with respect to the start of the discharge, densities smaller than the peak value can be reached. %
    The lowest density for which experiments were conducted is $\SI{0.5e14}{cm^{-3}}$. %
    Figure~\ref{fig:DPS-cams} a) shows the light signal recorded by the two cameras for an argon plasma at a pressure of 24\,Pa, a peak current of 500\,A and 80\,\textmu s after the start of the discharge. %
    The plasma density for these parameters was $\SI{4.8e14}{cm^{-3}}$~\cite{Carolina_DPS}. %
    The left-most and right-most images were obtained by moving camera 1 from one location to the other. %
    The images are corrected as described in Section~\ref{sec:DPS-setup}. %
    The regions where the images overlap are marked by red rectangles. %
    The profiles of the light captured along the plasma in Fig.~\ref{fig:DPS-cams} b) were obtained by summing counts vertically over a region 75 pixels wide, fully containing the discharge tube ($\sim$ 37 pixels wide). %
    Profiles for eight events are plotted on top of each other. %
    The narrow width of the resulting line shows the repeatability of the discharge and measurement. %
    A single signal varies between $\pm 2.2 \%$ (minimum to maximum) over a short section, while the mean of many signals over that same section varies by only $\pm 0.2 \%$. %
    On the same figure we also plot the profiles of uncorrected images in lighter colors. %
    The profiles of uncorrected images bend down towards their edges. %
    This is mainly caused by vignetting, but also by barrel distortion. %
    The geometry of observing the plasma at an angle has an opposite effect, and partly counteracts the downward bending to some extent. %
    The corrected profiles show a variation in signal amplitude of about 12\% (minimum to maximum) over the length of the plasma. %
    At this time, it is unclear whether this non-uniformity is introduced by limitations in the method for acquiring/correcting the images, or due to the amount of light emitted by the plasma. %
    This warrants further investigation into the uniformity of the plasma, both in shape (e.g. width of the discharge tube) and in density. %

    \subsection{Vapor Plasma}
\begin{figure}[!b]
        \centering
        \includegraphics[width=1\textwidth]{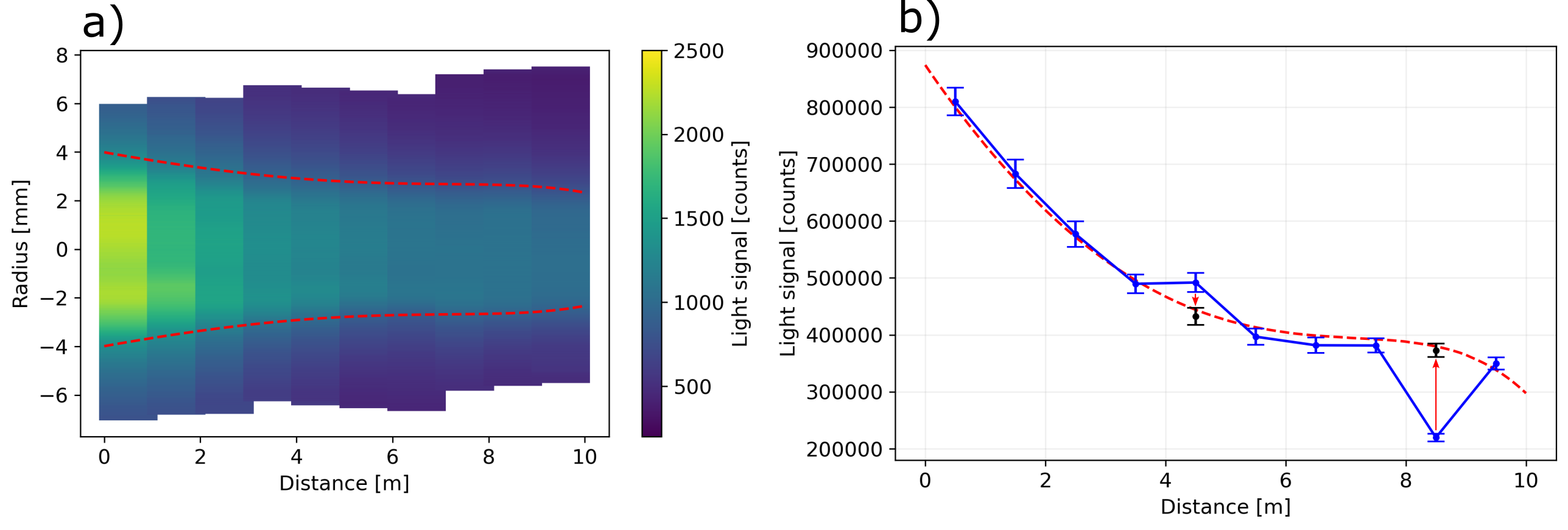}
        \caption{(a) Counts of the light captured by cameras summed in horizontal direction over the ROI at each view port position. %
        For presentation purposes, the slices are aligned vertically so that the center of the plasma column is at 0\,mm for all slices. %
        Images at 4.5 and 8.5\,m corrected according to Fig. b), see text.
        Red dotted lines: predicted radius of the plasma column, with values obtained from the fit in (b). %
        (b) Summed pixel counts of the ROI at each view port position (blue symbols).
        Error bars represent the standard deviation of the dataset,
        consisting of 14 events with a laser pulse energy between 89 and 90\,mJ and plasma density $n_{pe} = (1.957 \pm 0.005) \times 10^{14} $ $\mathrm{cm}^{-3}$. %
        Red dotted line: fit of the model for the area of the plasma column along the plasma. %
        Initial laser energy in the model 90 mJ. %
        Correction factors applied for measurements at 4.5\,m and 8.5\,m to agree with fitted model (blue to black symbols). 
        }
        \label{fig:VAPS-envelope}
    \end{figure}

    In the VPS, the plasma is created in a rubidium vapor by an ionizing laser pulse~\cite{oz2014VAPS}. %
    Plasma densities between $\SI{1e14}{cm^{-3}}$ and $\SI{10e14}{cm^{-3}}$ can be reached. %
    The intensity of the laser pulse ($\sim \SI{10}{TW/cm^{2}}$) is sufficient to ionize the first electron of the rubidium atoms ($e\phi_{1} = \SI{4.18}{eV}$) through field ionization, but not the second one ($e\phi_{2} = \SI{27.29}{eV}$). %
    A plasma column where 100\% of rubidium atoms are singly ionized~\cite{adli2019experimental}, over a radius that depends on the laser pulse radius and intensity, is created. %
    Figure~\ref{fig:VAPS-envelope} a) shows signals obtained from the ten cameras along the source by summing the counts in the ROI in horizontal direction (see Fig.~\ref{fig:VAPS-setup}) for a plasma with $n_{pe} = (1.957 \pm 0.005) \times 10^{14} $ $\mathrm{cm}^{-3}$. %
    The width of the plasma and the intensity of the light changes with distance from the plasma entrance. %
    This is because the transverse size of the laser pulse changes as it propagates and its energy decreases as it ionizes the vapor. %
    Figure~\ref{fig:VAPS-envelope} b) shows the sum of the counts for each camera (blue symbols). %
    Eight out of ten measurement points follow a decreasing trend. %
    The amount of light emitted through recombination is proportional to the number of electron-ion pairs created at the locations along the plasma. %
    Since the rubidium vapor density $n_{rb}$ is uniform to within 0.2\%~\cite{oz2014VAPS}, the plasma density within the 100\% ionized region is also uniform. %
    Therefore, the amount of light emitted should scale linearly with the volume of the plasma column %
    in the observed slices of the plasma. %
    The transverse area of the plasma can be estimated by a simple model, similar to that shown in reference~\cite{oz2014VAPS}.
    We first describe the model and then show that this model fits the experimental data and explains the trend observed by the cameras. %
    
    The plasma is created using a Ti:Sapphire laser pulse~\cite{awake_terrawatt_laser} with a center wavelength of $\lambda = \SI{780}{nm}$. %
    The propagation of the envelope of the laser pulse in vacuum is described by 
    that of an approximately Gaussian beam. %
    The envelope is characterized by the waist radius $w_0$ and position $z_0$, Rayleigh length $z_R = \frac{\pi w_0}{\lambda^2}$, and $M^2$ factor. %
    The beam size along the axis of propagation z is given by: 
    \begin{equation}
        w(z) = w_0 \sqrt{1 + M^4\left( \frac{z-z_0}{z_R} \right)^2}
    \end{equation}
    \noindent where $z = 0$ is located at the plasma entrance. %
    The measured beam parameters in vacuum for the laser beam are $z_0 = \SI{5.48}{m}$, $w_0 = \SI{1.65}{mm}$ ($2\sigma$), $z_R = \SI{10.97}{m}$ and $M^2 = 2$. The laser pulse energy varies between 80 and 120\,mJ ($\sim$ 90\,mJ for the results presented here)%
    and the pulse duration is $t_{laser} \approx \SI{120}{fs}$. %
    For field ionization, the intensity of the laser pulse has to be larger than the threshold for field ionization $I_{th}$. %
    The intensity of the laser pulse at position z and distance to the beam axis r is approximately given by:
    \begin{equation}
        I(r,z) = I_0 \left(\frac{w_0}{w(z)}\right)^2 e^{\frac{-2r^2}{w(z)^2}}.
    \end{equation}
    \noindent with the peak intensity $I_0$. %
    The boundary of the plasma column is the location where the intensity of the laser pulse is equal to the threshold value for field ionization. %
    The transverse area within the boundary is then given by: 
    \begin{equation}
    \begin{aligned}
    A_{plasma}(z) & = -\pi \frac{w(z)^2}{2}\ln{\left( \frac{I_{th}}{I_0} \left(\frac{w(z)}{w_0}\right)^2 \right)}
    \end{aligned}
    \end{equation}
    Because the peak intensity scales with the laser pulse energy, we calculate the energy lost due to ionization of the plasma column for a slice of thickness $d_{step}$ and area $A_{plasma}(z)$ as: %
    \begin{equation}
    E_{loss}(z) = A_{plasma}(z) \cdot d_{step} \cdot e\phi_1 \cdot n_{rb}
    \end{equation}
    For the next slice at $z+d_{step}$, we calculate the area of the plasma column (Eq. 3) with the lower peak electric field %
    $I_{0,step} = 2\frac{E_{laser} - E_{loss}}{\pi w_0^2t_{laser}}$. %
    We evolve the energy of the laser pulse, and thus the radius over which ionization occurs along the plasma, using this simple model. %
    We divide the plasma into small slices %
    of length $d_{step} = 1$\,cm, %
    and iterate this process. %
    This step size is sufficiently small, as the energy loss per step is small when compared to the laser energy ($E_{loss} \ll E_{laser}$). %
    We calculate the energy lost through ionization of the vapor slice (Eq. 4), and subtract it from the energy of the pulse for the next slice, taking into account the evolution of the beam size in vacuum (Eq. 1). %

    We fit the result of this calculation to the measured data (Fig.~\ref{fig:VAPS-envelope}~b), red dotted line) with two free parameters. %
    The first one is the %
    threshold for ionization $I_{th}$, which defines the extent of the boundaries of the plasma through the %
    intensity it takes to ionize the atoms. %
    The second one is a factor $a$ that we introduce to scale the energy lost at each step. %
    This factor accounts for additional energy loss due to atomic excitation and partial ionization outside of the fully ionized plasma column, and to kinetic energy that electrons gain from the pulse. %
    The free parameters obtained from the fit are %
    $I_{th} = \SI{0.42}{TW/cm^2}$
    and $a = 6.8$. %
    The theoretical value for the threshold of field ionization of rubidium is %
    $I_{th} = \SI{1.7}{TW/cm^2}$~\cite{rb_threshold}.
    It is not surprising that the value we obtain for %
    $I_{th}$
    from the fit is lower than its expected value. %
    The spectrum of the laser pulse overlaps with atomic transitions from the ground state and first excited state of the rubidium atom. %
    Therefore, resonant excitation to the first (5P$_{1/2}$) and second (5D$_{3/2}$,5D$_{5/2}$) upper states decrease the threshold for ionization that occurs from these excited states~\cite{demeter2019propagation}. %
    The value we obtain for $a$ is large (1 for ionization only). %
    However, it is commensurate with simulations and experimental results~\cite{PhysRevA_104_033506} showing that the energy that the laser pulse loses is several times larger than the energy it would take for only ionization of the plasma channel. %

    Figure~\ref{fig:VAPS-envelope}~b) shows that eight out of the ten measurement points agree with the fitted model. %
    For the two measurements that do not agree, we attribute the differences to the calibration and alignment of the cameras, as well as variations in the setup. 
    Therefore, we determine correction factors for these two measurement points so that they agree with the values expected from the model. %
    In Fig.~\ref{fig:VAPS-envelope} a), we plot the boundaries of the plasma channel (red dotted lines), as predicted by the model. %
    We observe that most of the recorded light is indeed emitted from within these boundaries (correction factors for images at 4.5\,m and 8.5\,m applied). %
    The agreement between the model and the measured data shown in Fig.~\ref{fig:VAPS-envelope} indicates that the camera signals record the energy deposited by the laser pulse in the plasma %
    at the ten locations of the observed volume (e.g. red rectangle in Fig.~\ref{fig:VAPS-setup} b)-c)). %
    
\begin{figure}[!b]
        \centering
        \includegraphics[width=1\textwidth]{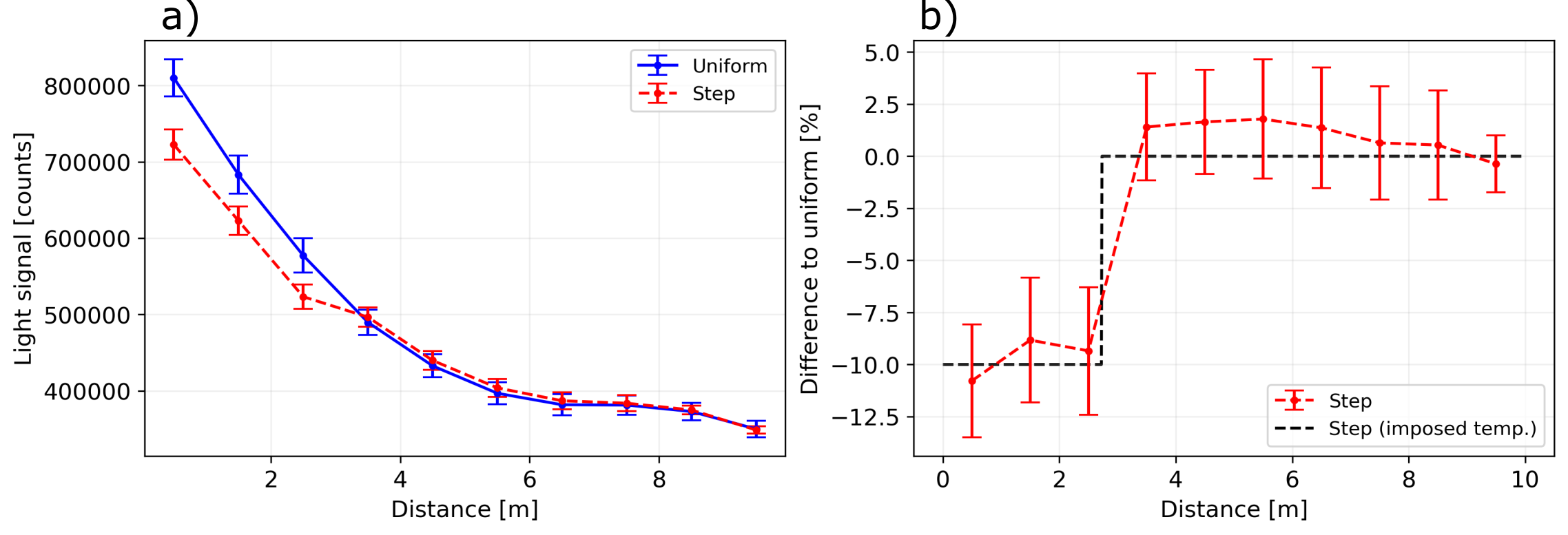}
        \caption{(a) Pixel counts sum from the ROI at each viewport position with a uniform plasma (blue symbols, same as Fig. \ref{fig:VAPS-envelope} b)), and with a plasma with a density step (red symbols). Correction factors applied (see. Fig. \ref{fig:VAPS-envelope} b)). %
        Error bars standard deviation of the dataset, 14 (blue) and 8 (red) events, laser pulse energy 89-90\,mJ. %
        (b) Relative difference of the signals of Fig.~a). %
        Black dashed line: step function representing the ideal step (temperature imposed on VPS segments).}
        \label{fig:VAPS-step}
    \end{figure}

    Since the amount of energy deposited depends on the plasma density (see Eq. 4), which is equal to the vapor density in our case, we can use the light diagnostic to observe different density profiles along the source. %
    For example when imposing a step in the density while keeping all other parameters the same. %
    Such a step may be used to prevent the amplitude of the wakefields to decrease after their saturation point~\cite{lotov2015physics}. %
    By increasing the temperature of the VPS by 10\% over the first 2.735\,m of the VPS (temperature step), we lower the rubidium vapor density by the same relative amount (-10\%), according to the ideal gas law. %
    White light interferometers near the plasma entrance and exit of the source measure the rubidium vapor densities~\cite{batsch2018interferometer} as $n_{rb} = (1.793 \pm 0.004) \times 10^{14} $ $\mathrm{cm}^{-3}$ and $ (1.962 \pm 0.004) \times 10^{14}$ $\mathrm{cm}^{-3}$ ($\pm$ standard deviation over 12 measurements), respectively. %
    Thus, the relative height of the step in vapor density we measure is $(-9.43 \pm 0.45) \%$. %
    Figure~\ref{fig:VAPS-step} a) shows the measurements of the light emitted with the uniform vapor density (blue symbols) and with the density step (red symbols). The two measurements are independent of each other. The correction factors determined in Fig. \ref{fig:VAPS-envelope} b) are also applied here. %
    For the first three measurement points ($z$ = 0.5, 1.5, 2.5\,m), where the vapor density is lower, less light is observed. %
    The relative difference between the measurement with and without the step (Fig. \ref{fig:VAPS-step} b)) shows that the signal with a density step (red symbols) is approximately 10\% lower for the first three measurements ($(-9.7 \pm 1.7) \%$) and is in agreement with the expected values for the imposed step (black dashed line). The other seven measurements are not significantly different to the measurements at uniform plasma density ($(0.6 \pm 0.9) \%$). The relative height of the step in light emission of the plasma we measure is $(10.3 \pm 2.6) \%$.
    
    The fact that the measurement at 2.5\,m, i.e. 23.5\,cm in front of the point where the (temperature) step is imposed, is not significantly different to those at 0.5 and 1.5\,m indicates that, as expected~\cite{plyushchev2017rubidium}, the half width of the step in density extends over less than 23.5\,cm. %
    The uncertainty (standard deviation) on the plasma light measurements is between $\pm 1.4 \%$ and $\pm 3 \%$, and is significantly larger than the uncertainty on the measured rubidium vapor density using the interferometers. %
    Nevertheless, using the light diagnostic enables us to sample more points along the plasma source than with white light interferometry, and to verify the location and relative height of the density step, and put an upper bound on its sharpness. %

\section{Conclusion}

Utilizing the light emitted by plasma as a diagnostic for wakefield accelerators has been demonstrated with short bunches as driver~\cite{SLAC-plasmalight,boulton2022longitudinally, zhang2024correlations}.
For a plasma wakefield accelerator based on self-modulation of long drivers, where the wakefield amplitude evolves along the plasma, the diagnostic has to be able to record the light at many points along the plasma.  %
In this paper, we presented the implementation of light diagnostics for two AWAKE plasma sources.
We utilized CMOS cameras 
to study the light emitted along ten meters of plasma. %
We discussed the setup and relative calibration of the camera systems. %
Images of the discharge plasma with wide angle lenses were corrected for distortion, vignetting and the observation angle of the discharge tube. %
For the vapor plasma source, a model of the expected amount of light from the laser-ionized plasma, was fitted to the ten camera signals.
The agreement between the model and the experimental data indicates that the amount of light observed is proportional to the energy locally deposited in the plasma (no wakefields in this case). %
We can therefore, in principle, use this diagnostic to investigate the development of self-modulation along the plasma, assuming the energy deposited and dissipated at each location is proportional the energy stored in the wakefields. %
We exploited the proportionality between emitted light and plasma density in the vapor source to confirm for the first time the height, position and sharpness ($\leq$~23.5\,cm) of an imposed density step.

\section{Acknowledgments}
\noindent The authors thank Dr. Eugenio Senes for help with the installation and calibration of the DPS cameras.

\bibliography{bib}

\end{document}